\documentclass[twocolumn,snoshowpacs,epl,preprintnumbers,superscriptaddress,amsmath,amssymb,floatfix]
{revtex4}
\usepackage{graphicx}
\usepackage{color}
\usepackage{placeins}
\usepackage{tikz}
\usepackage{mathtools}
\usepackage{amsthm}
\usepackage{romannum}
\usepackage{xkeyval,xcolor}
\usepackage{amsmath}
\usepackage{caption}
\usepackage{subcaption}
\usepackage{ragged2e}
\usepackage{epstopdf}
\DeclareCaptionJustification{justified}{\justifying}
\begin{document}

\newcommand{\er}{Erd\H{o}s-R\'{e}nyi}
\newcommand{\red}{\color{red}\footnotesize}

\title{Two phase transitions in modular multiplex networks}

\author{Yael Kfir-Cohen}
\affiliation{Bar-Ilan University, Ramat Gan, Israel}
\author{Dana Ben Porath}
\affiliation{Bar-Ilan University, Ramat Gan, Israel}
\author{Bnaya Gross}
\affiliation{Northeastern University, Boston, MA, United States of America}
\author{Sergey Buldyrev}
\affiliation{Yeshiva University, New York, NY, United States of America}
\author{Shlomo Havlin} 
\affiliation{Bar-Ilan University, Ramat Gan, Israel}
\date{February 2026}

\begin{abstract}
    Modular networks, such as critical infrastructures, are often built from distinct, densely connected modules (e.g., cities) that are sparsely interconnected. When such networks are gradually and randomly disrupted under a percolation process, they undergo two critical phase transitions. The first transition occurs when modules become isolated from one another, while the second corresponds to the collapse of the entire network, including the internal connectivity of the modules. Here, we study these phase transitions in modular multiplex networks and compare them with those observed in single-layer modular networks. We focus on models in which the modules are arranged and connected either as a Random Regular network or as a two-dimensional square lattice. We show here that these systems exhibit diverse transition behaviors, with some transitions occurring continuously and others abruptly; notably, one realistic model could display two distinct first-order transitions in the same system. For the modular Random Regular multiplex, we further characterize the spatial transition through its scaling behavior, revealing signatures of a mixed-order phase transitions. In addition, we analytically determine the critical threshold at which modules become disconnected. Our results highlight the crucial role of modular organization and the critical role of interdependence in shaping network vulnerabilities under failures.
\end{abstract}
  \maketitle

\section{INTRODUCTION}
Network theory has been a central tool for understanding complex systems in recent decades.
It became clear over the years that the stability and durability of such systems depend greatly on their structure.
% modular:
Many complex systems are built in a modular structure where the nodes are highly connected inside the modules and have less connections between the modules \cite{BAGROW2015,shekhtman_2015,KfirCohen2022,bianconi2018multilayer}. For example, urban infrastructures in a country, are highly connected inside the cities and also have connections between the cities \cite{eriksen2003modularity,guimer2005}.
Other examples of modular systems are the brain \cite{Meunier2010,Stam2010,Morone2017} and social circles \cite{CHEN200711,Rybski2011}.
%  multiplex:
In addition, real-world systems are composed of different network layers, and accordingly, they have a set of connections within each network and dependency links between the networks. Examples include, Electricity and Internet networks and the dependency between them \cite{buldyrev-nature2010,boccaletti-physicsreports2014, Dong2021,hu-prx2014,Gross_2020}.

Here, we investigate the intriguing emergence of two distinct phase transitions in modular multiplex networks, studied within the framework of percolation processes \cite{Coniglio_1977,coniglio-jphysicsa1982,Sokolov_1986,staufferaharony,bunde1991fractals,morone-nature2015,mureddu-scientificreports2016}. We consider two models: (a) a multiplex modular network in which inter-module links are restricted to the four nearest neighboring modules, analogous to the adjacency pattern of a two-dimensional lattice; and (b) a multiplex modular network in which each module is connected to a fixed number of other modules without spatial proximity constraints, as in a Random Regular (RR) network. In addition, we compare the resulting transitions in these multiplex systems with the phase transitions reported by Gross \textit{et al.} \cite{Gross2020} for a single-layer modular network arranged on a $2D$ lattice.

Our simulations reveal that both models exhibit two distinct phase transitions: the first, in which the modules disconnect from one another, and the second, where the modules collapse, which occurs near the typical phase transition of Erdős–Rényi (ER) multiplex networks, since each module constructed as a small ER network.
The nature of these transitions varies across the different models, with some appearing continuous while others are abrupt. Moreover, we developed a theoretical framework that accurately predicts the critical point at which the first transition in which the modules become disconnected from each other. Finally, we investigate the critical exponents and the underlying mechanism associated with the newly abrupt phase transition, and show that both align with those previously observed in interdependent networks and in physical networks \cite{buldyrev-nature2010, Dong2021,Bonamassa2023}.

\begin{figure}[ht]
		\centering
    % \subfloat[]
    {\includegraphics[width=\linewidth]{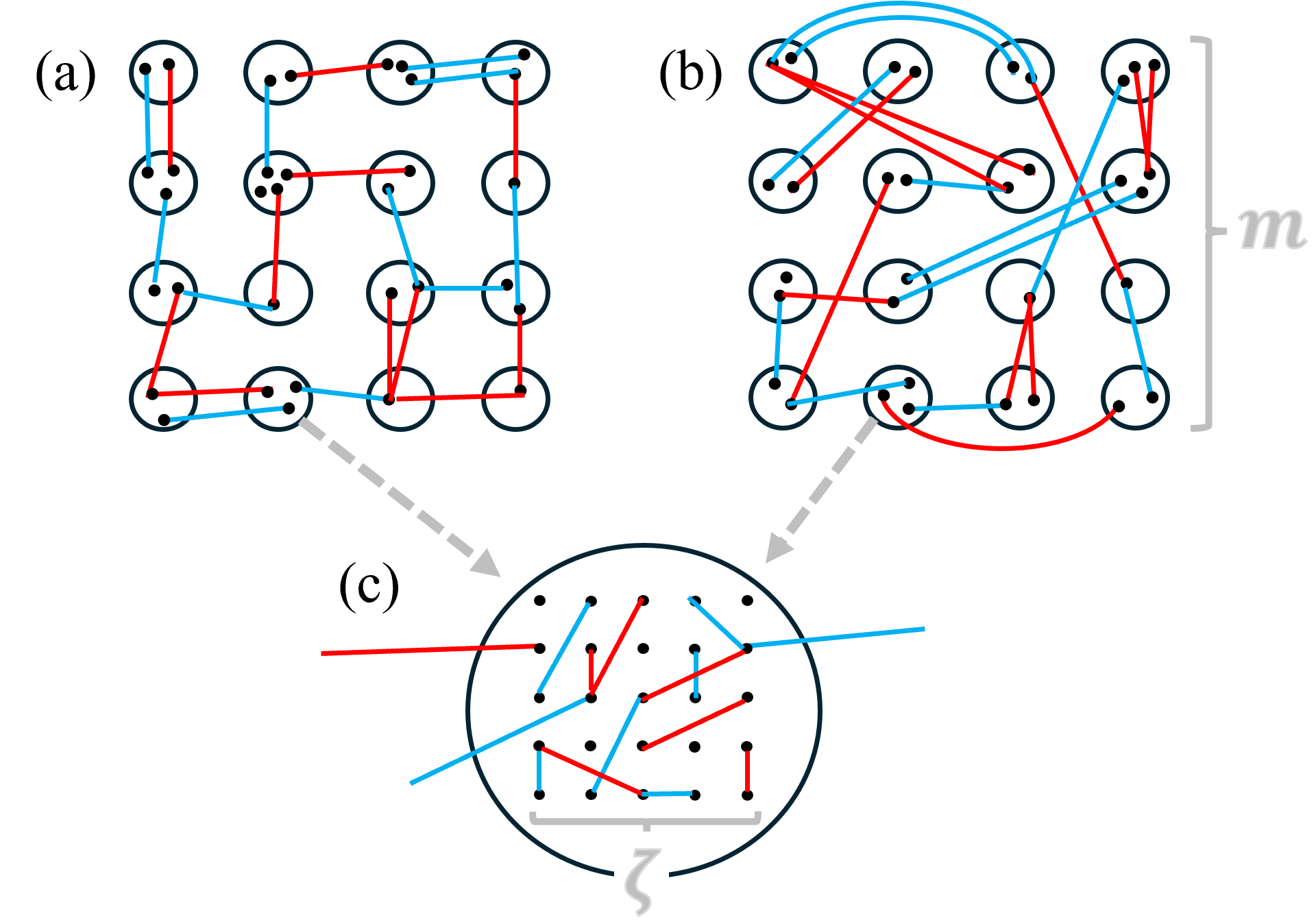}}
% {\includegraphics[width=0.3\linewidth]{1a model multiplex modular lattice.png}}
% \hspace{9pt}
% \subfloat[]
% {\includegraphics[width=0.3\linewidth]{1b model multiplex modular RR.png}}
% \hspace{9pt}
%   \subfloat[]
% {\includegraphics[width=0.3\linewidth]{1c model multiplexER.png}}\\
    \caption{
        \textbf{A schematic representation of our models.} 
        The red and blue lines represent the links in the first and second layer in the multiplex.
        Each circle represents a module and each dot represents a node.
        (a) and (b) Modular multiplexes, while drawing only inter-nodes and inter-links. 
        In (a) the inter-links connect each module to some of its $4$ closet modules in the lattice of modules, while in (b) the interlinks connect any module to any $D$ modules (Here, $D = 1$) which may be at any distance.
        (c) Demonstrating a close-up of a module in both models (a) and (b), with interlinks and intra-links which are distributed Poissonian as an $ER$ network.
    }
\label{fig:Demonstration2}
\end{figure}

\section{Model}
Here, we study two models of multiplex modular networks and compare them to a single-layer modular network model. In all the models, the network is composed of modules while each module includes $\zeta \times \zeta$ nodes and the whole network includes $L \times L$ nodes. Thus, the network is built from $m \times m$ modules while $m=L/\zeta$.
Every node in the network may have links to nodes that are part of its module and to nodes that are part of other modules.
The intra-degree of the intra-links within the modules is defined as $k_{intra}$ and the inter-degree of the inter-links between neighboring modules is defined as $k_{inter}$. The degree distributions are Poissonian, implying that each module effectively behaves as a small Erdős–Rényi (ER) network.
We define $Q$ as the averaged degree of a module connected to other modules and it is equal to $k_{inter}\cdot\zeta^2$.
 
In our new modular multiplex networks (models (a) and (b)), each node participates in two distinct layers of connectivity, i.e. it has two sets of links, and is considered functional only if it belongs to the giant component in both layers, see Fig \ref{fig:Demonstration2}(c).
Similar to the structure studied for the single-layer modular model \cite{Gross2020}, in model (a), each layer consists of modules organized in a two-dimensional square lattice structure.
Each node in the network may have links to nodes within its own module and to nodes in the four nearest neighboring modules in the lattice structure, see Fig \ref{fig:Demonstration2}(a). 
In model (b), each layer comprises modules organized in a Random Regular ($RR$) network.
We fix the number of modules to which each module is connected and denote this number by $D$. Accordingly, nodes may form links both within their own module (intra-links) and to nodes in $D$ other modules (inter-links), which are not required to be spatially or topologically neighboring, see Fig \ref{fig:Demonstration2}(b). 

\section{Results}
\begin{figure}[ht]
		\centering
\includegraphics[width=\linewidth]{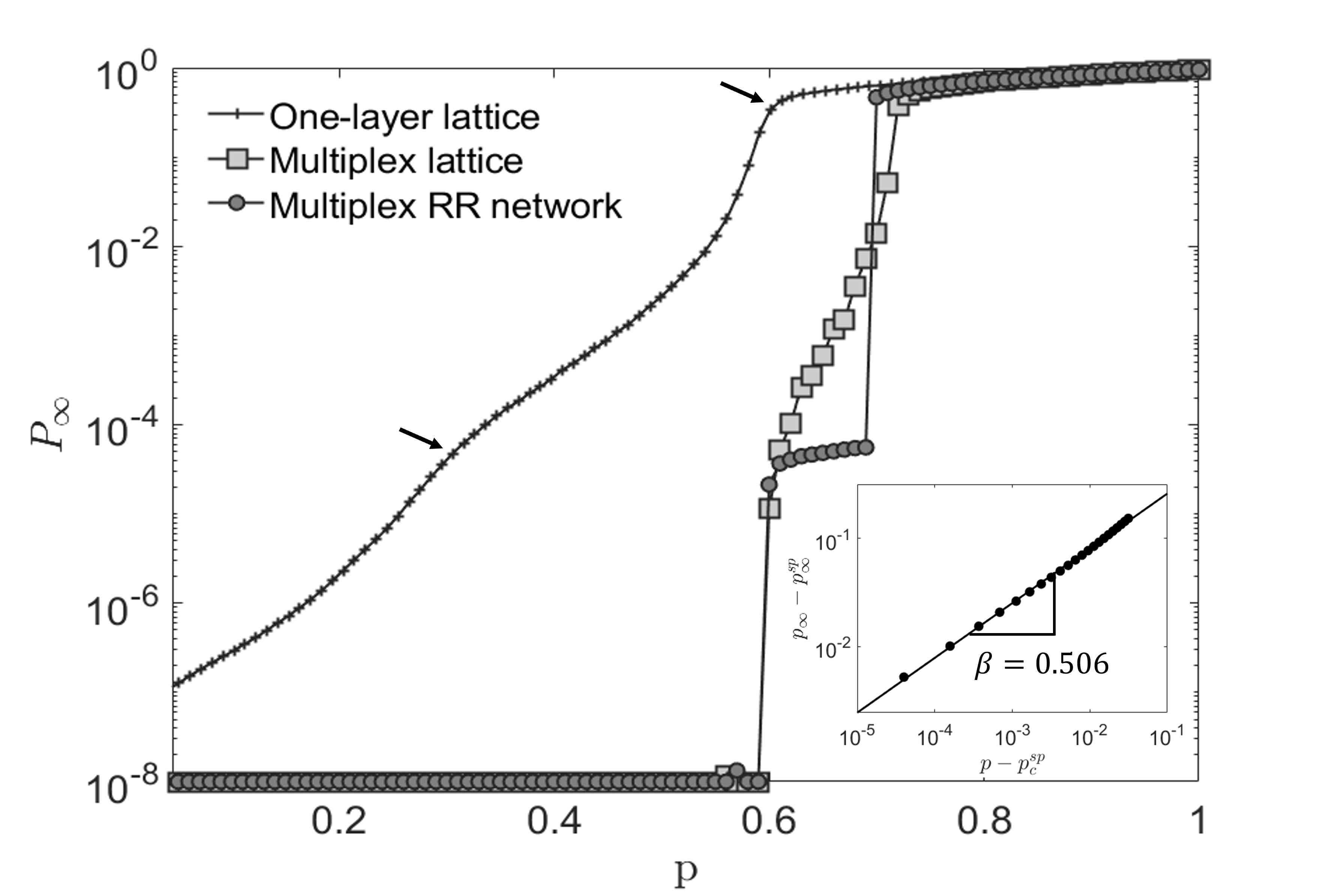}
    \caption{
        \textbf{Simulation results for the largest cluster $P_{\infty}$ as a function of $p$}. We compare three models: (i) a single-layer lattice model, which exhibits two continuous phase transitions; (ii) a multiplex lattice model, displaying one continuous and one abrupt (mixed-order) transition; and (iii) a multiplex random regular ($RR$) network, which exhibits two abrupt transitions. Notably, the lower transition point is identical for both multiplex models, as it corresponds to the fragmentation of an individual multiplex Erdős–Rényi ($ER$) module. \textbf{Inset:} The change of the giant component of the multiplex $RR$ network from its value at criticality as a function of the distance of $p$ from $p^{sp}_c$ in log log scale. From that we obtain $\beta=0.506$.
        Parameters used are $L=10^4$, $k_{intra} = 4$, $\zeta=100$, $k_{inter}= 10^{-3}$ and $Q=10$.}
    \label{fig:comparing3models}
\end{figure}

 \begin{figure}[ht]
		\centering
	\subfloat[]{\includegraphics[width=0.5\linewidth]{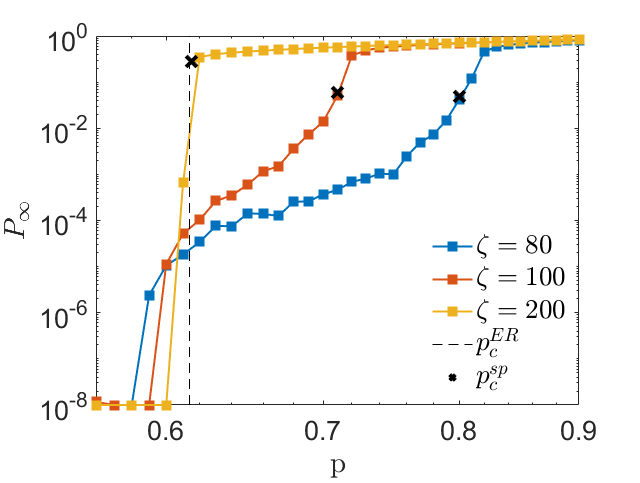}}
    \subfloat[]{\includegraphics[width=0.5\linewidth]{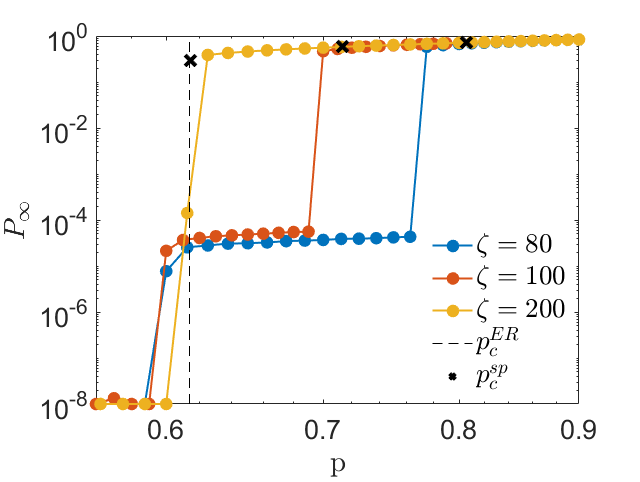}}\\
\caption{\textbf{Simulation results for the largest cluster $P_{\infty}$ in the two multiplex models as a function of $p$} for different values of $\zeta$ on semi log scale. Two distinct transitions are observed. The first (higher p) transition at $p^{sp}_{c}$ of the lattice. The second (lower p) transition occurs when the small $ER$ modules collapse at, $p^{ER}_{c}$. (a) Multiplex of a modular network where the modules are connected as in 2D lattice, and (b) multiplex of a modular network where the modules are connected as a $RR$ network. Parameters used are $L= 10^4$, $k_{intra} = 4$ and $k_{inter}= 10^{-3}$.}
		\label{fig:2transitions multiplex}
	\end{figure}

Here, we study the phase transitions observed in each model under a percolation process when removing a fraction $1-p$ of nodes randomly from the multiplex modular network.
In Fig. \ref{fig:comparing3models} we present in a log scale the behavior of the largest cluster size $P_\infty$ as a function of $p$ for our two multiplex models, compare between them and compare them to the one-layer model that have been studied earlier \cite{Gross2020}.
% In Fig. \ref{fig:comparing3models} we present the some specific values of $\zeta$ and fixed $k_{inter}$.
We observe that for fixed values of the systems, all three models demonstrate similar behavior of involving two phase transitions. However, in the one-layer modular lattice, both transitions are continuous, in the multiplex modular lattice, one is continuous while the other is abrupt. Surprisingly, in the multiplex modular RR network a new and unique phenomenon of two abrupt transitions appears.

In the three graphs, the upper transition corresponds to the network’s fragmentation into disconnected modules, reducing the giant component to the size of a single module, while the lower transition signals the collapse of the last remaining module.
Since every module is constructed as an $ER$ network, the lower transition behave similarly to the typical transition of $ER$ networks. It is continuous for the single-layer model and abrupt for the multi-layer models (a) and (b) due to the cascading process occurs between the layers which collapses the network abruptly.
We hypothesize and discover as seen in Fig. \ref{fig:2transitions multiplex} (a) and (b) that the critical probability threshold, $p_c^{ER}$, of the second transition approximates the typical threshold of an $ER$ multiplex network which is $2.4554/k_{intra}$ \cite{buldyrev-nature2010}.
In addition, it can be seen that when we increase $\zeta$, i.e. we increase the module's size, but save the inter-degree of any node fixed, the modules are highly tied and thus the system has only the typical $ER$ phase transition while the other phase transition disappears. 

The upper transition is influenced by the spatial structure of the inter-module connections (sp).
In model (a) (as well as in the one-layer model), the modules are arranged as nodes in a $2D$ square lattice, while in model (b), they are arranged as nodes in a RR network.
We assume that the network in model (a), and the network in model (b) break into individual modules in a manner similar to how nodes in a lattice and a RR network, respectively, become isolated under a bond percolation process.
The modules become isolated or small clusters of modules at the first upper transition, which corresponds to the transitions in the non-modular models where the nodes become small clusters.
This correspondence accounts for the behavior observed in Fig. \ref{fig:comparing3models}. 
% The spatial transitions in model (a) and the one-layer model are second-order and continuous, since the connectivity links are short-range, much like the phase transition in a non-modular lattice, whereas the spatial transition in model (b) is a first-order and abrupt transition, characteristic of a RR multi-layer phase transition.
Since all dependency links in our multiplex models are short-range, (within the same node) and the connectivity links between modules in model (a) are also short-range, the spatial transition in model (a) is continuous and second-order, as in non-modular lattices. In contrast, the long-range connectivity links between modules in model (b) give rise to an abrupt, first-order spatial transition, characteristic of RR multiplex networks.
Additionally, we use this understanding to develop a new analytical theory for the spatial transition threshold value.
Our method compares the probability that there no interlinks between two modules in our modular models, with $1-b$, the probability of not having a link between two nodes in one of the layers of the respective non-modular multiplex models.
Since we are interested in the spatial transition threshold, we compare these probabilities at the critical threshold $b_c$.
% We mark the probability for a link between two nodes in the non-modular models by $b$, then the probability there isi no link is $1-b$.
We first derive the mathematical expression for $p_c^{sp}$, which depends on the critical threshold $b_c$ of the non-modular networks. We then compute $b_c$ for the non-modular multiplex lattice and cite the known result for the multiplex RR network. Substituting both values of $b_c$ into the expression of 
$p_c^{sp}$, we validate our results for models (a) and (b), respectively.

\begin{figure}[ht]
	\centering
	\subfloat[]{\includegraphics[width=0.5\linewidth]{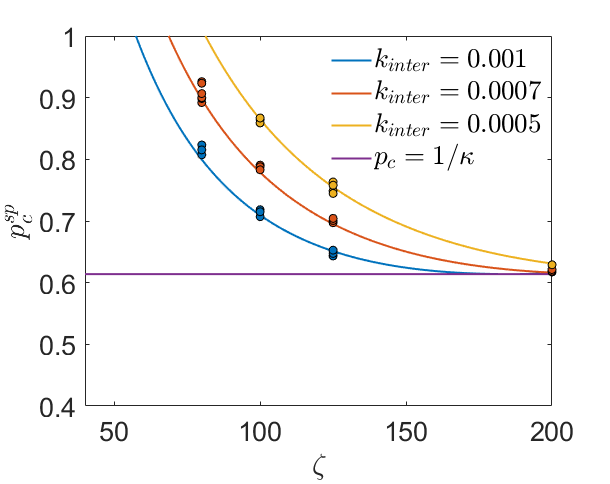}}
	\subfloat[]{\includegraphics[width=0.5\linewidth]{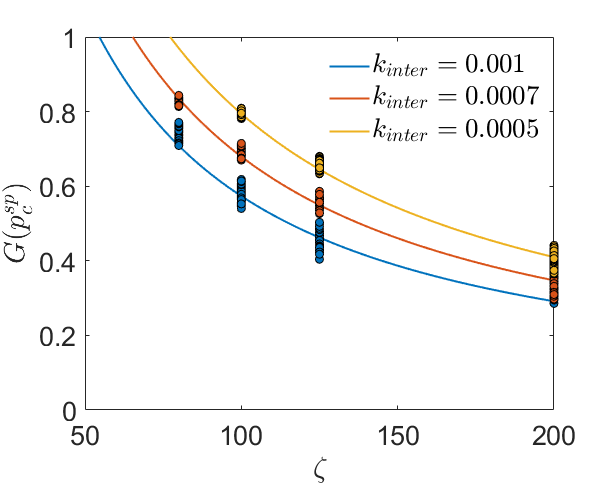}}\\
    \caption{
        \textbf{Multiplex modular lattice} (model (a)). 
        \textbf{(a)} $G(p^{sp}_{c})$ and \textbf{(b)} $p^{sp}_{c}$ as a function of $\zeta$ for various values of $k_{inter}$. The circles represent simulation results and the lines are the theory obtained from (a) equation (3) and (b) equation (4). In the limit of $\zeta \rightarrow \infty$ the system approaches a single ER network and $p^{sp}_c$ approaches $p^{ER}_{c}$. Parameters used are $L= 10^4$ and $k_{intra} = 4$.}
	\FloatBarrier
    \label{fig:giant and pc}
\end{figure}

In our modular models, we consider $P_k(Q)$ as the binomial distribution that out of $Q$ interlinks originating from a module, $k$ interlinks connecting to nodes of another specific module.
The probability that an interlink emanates from a given module with $D$ connected modules connects to one of them is $1/D$.
Therefore, $P_k(Q)=(1/D)^k(1-1/D)^{Q-k}C_Q^k$.
In addition, $G$ is the giant component of a single module, thus $1-G^2$ is the probability that an interlink does not reach, at both ends, nodes that participate in their local giant components.
Then, the analytical expression for the probability that two modules lack a single interlink connecting their local giant components is:
\begin{equation}
\label{eq:expression}
\begin{gathered}
\sum_k P_k(Q)(1-G^2)^k=[1-\frac{G^2}{D}]^Q.
\end{gathered}
\end{equation}
Thus, comparing $1-b_c$ with this analytical expression we obtain $G$ at the spatial transition:
\begin{equation}
\label{eq:G}
\begin{gathered}
G^{sp}_c=\sqrt{D(1-(1-b_c)^{1/Q})}.
\end{gathered}
\end{equation}
Since every module in our models is an $ER$ multiplex network, it is known that the giant component of a module maintains $G=p(1-e^{-k_{intra}G})^2$ \cite{gao-pre2012} and thus:
\begin{equation}
\label{eq:pc}
\begin{gathered}
p=\frac{G}{(1-e^{-k_{intra}G})^2}.
\end{gathered}
\end{equation}
By substituting the expression of $G^{sp}_c$ from Eq. \ref{eq:G} we derive:
\begin{equation}
\label{eq:pc_sp}
\begin{gathered}
p_c^{sp}=\frac{\sqrt{D(1-(1-b_c)^{1/Q})}}{(1-e^{-k_{intra}\sqrt{D(1-(1-b_c)^{1/Q})}})^2}.
\end{gathered}
\end{equation}
For the spatial transition of model (a), we focus on the percolation threshold of a multiplex two-dimensional ($2D$) square lattice under bond percolation. To this end, we simulate a multiplex lattice composed of two identical layers of links, denoted A and B. A fraction $1-b$ of bonds is randomly removed from each layer. This bond removal initiates a percolation process in the system, as any node disconnected from the giant connected component of one or more layers is deemed non-functional. We find that the system collapses when the bond existence probability reaches $b=0.5767$. 
To achieve $p_c^{sp}$ we substitute this value at Eq. \ref{eq:pc_sp} with $D=4$, the number of connected modules of a module in model (a).
The $p_c^{sp}$ derived from this theoretical approach shows excellent agreement with our simulations, as shown in Fig. \ref{fig:2transitions multiplex}(a).
Further confirmation of our new theory comes from its strong alignment with simulations of the giant component of a single module, $G$, and of the spatial threshold, $p_c^{sp}$ for different $\zeta$ values, as shown in Fig. \ref{fig:giant and pc}(a) and \ref{fig:giant and pc}(b).

For the spatial transition of model (b) we use the critical threshold of $RR$ multiplex network which was found by Gao et al \cite{gao-pre2012} in Eq. $(58)$:
\begin{equation}
\label{eq:pcRR}
\begin{gathered}
\frac{1}{p_t} = \max_{0 \leq z \leq 1} [\frac{(1-z^{\kappa-1})(1-z^\kappa)}{(1-z)}] = R(z),
\end{gathered}
\end{equation}
where $p$ is the critical threshold and $\kappa$ is the $RR$ degree, i.e. the fixed number of nodes each node connects.
This model is consistent with our modular $RR$ framework when we identify the $RR$ degree $\kappa$ with the module degree $D$. Substituting the corresponding value of $p_t$ for $b_c$ in Eq. \ref{eq:pc_sp} for a given $D$, we obtain the analytical prediction for $p_c^{sp}$. Fig. \ref{fig:2transitions multiplex}(b) shows excellent agreement between this analytical threshold and the simulation results.

\begin{figure}[ht]
		\centering
  \subfloat[]{\includegraphics[width=0.5\linewidth]{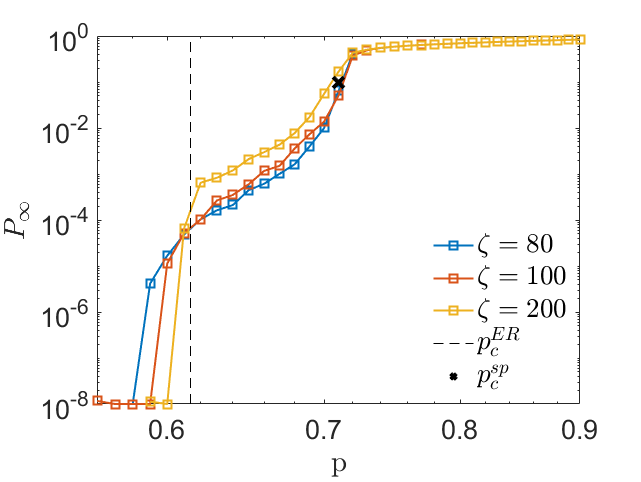}}
  \subfloat[]{\includegraphics[width=0.5\linewidth]{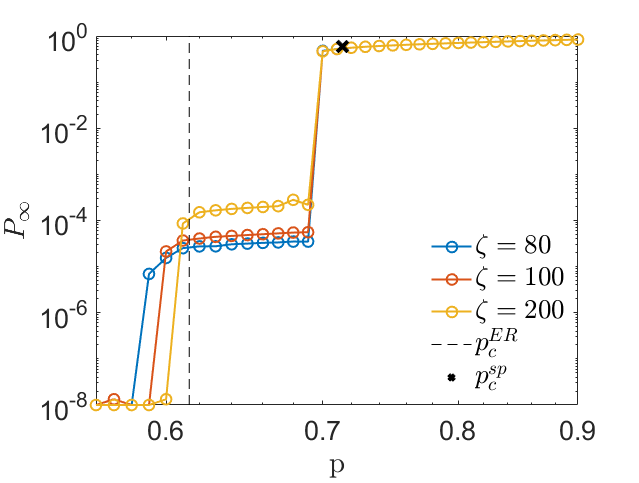}}\\
		\caption{{\bf Simulations of the largest cluster $P_{\infty}$ as a function of $p$} for different values of $\zeta$ on semi log scale with a fixed module degree, {\bf $Q=10=k_{inter}\cdot \zeta^2$}. Two distinct transitions are observed. The first (higher) transition at $p^{sp}_{c}$ of the lattice. The second (lower) transition occurs when the small $ER$ modules break apart, $p^{ER}_{c}$. (a) multiplex of a modular network where the models are connected as in 2D lattice, and (b) multiplex of a modular network where the models are connected as in a Random Regular network.
        Parameters used are $L= 10^4$ and $k_{intra} = 4$.
        }
		\label{fig:2transitions multiplex Q10}
	\end{figure}

In Fig. \ref{fig:2transitions multiplex Q10}(a) and (b) we present the same quantities shown in Fig \ref{fig:2transitions multiplex}(a) and (b) but for fixed average module degree, $Q$, rather than the average inter-degree of a node, $k_{inter}$.
Under this condition, we observe that the spatial transition threshold remains constant across all values of  $\zeta$ for all models, indicating that it does not depend on the module size. 
These results are fully consistent with the fact that Eqs. 2 and 3 depend only on the intra-degree of a node, the module degree Q, and the number of neighboring modules D.

\begin{figure}[ht]
	\centering
    {\includegraphics[width=\linewidth]{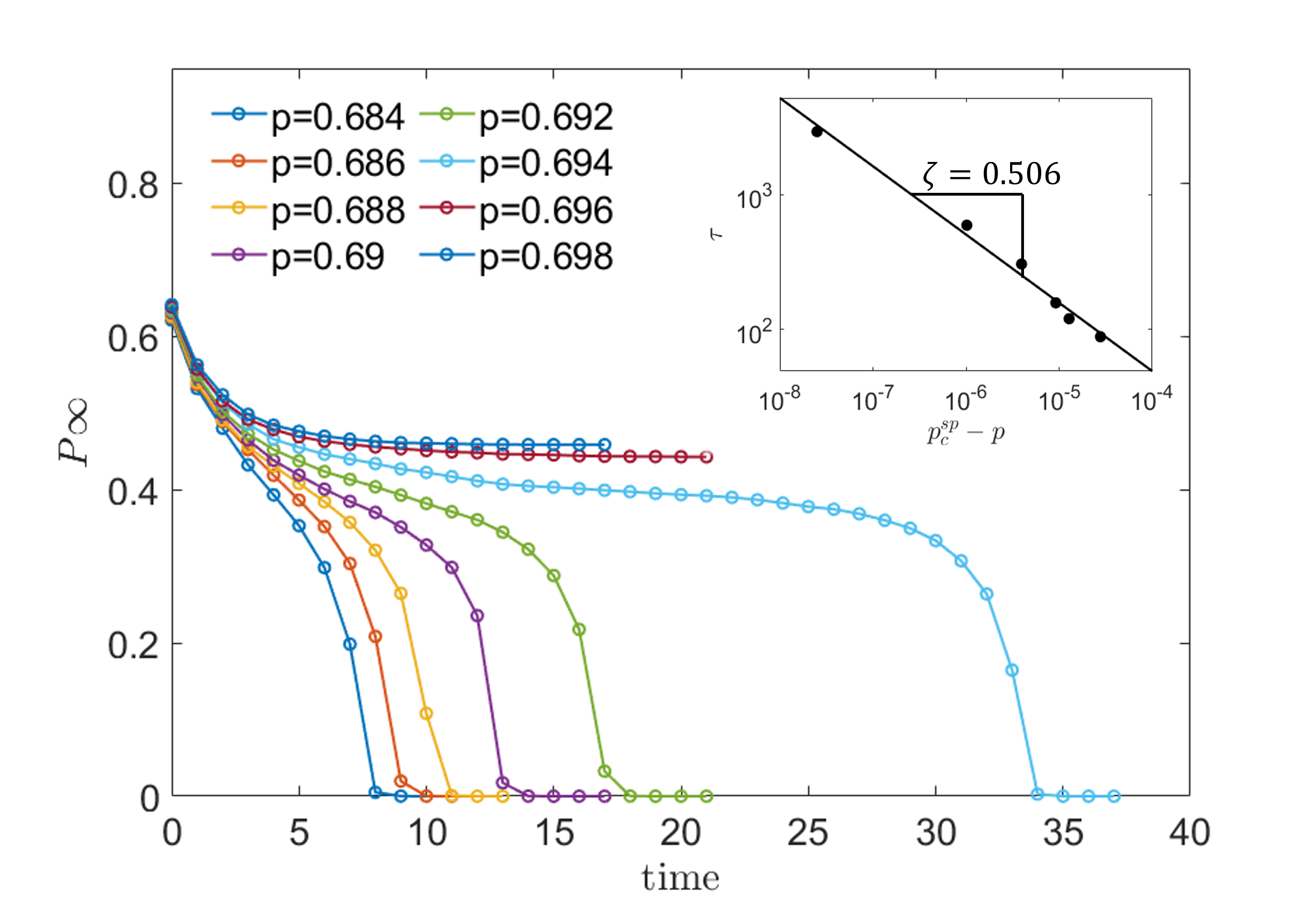}}\\
    \caption{
        \textbf{Multiplex modular $RR$ network} (model (b)). 
        The duration of the failure cascade, $\tau$, near and at the critical threshold of the spatial transition. At $p_c$ appears the longest plateau.
        \textbf{Inset:} The plateau duration as a function of $p_c-p$ in log log scale gives the critical exponent $\zeta=0.506$.
        Parameters used are $L= 10^4$, $k_{intra} = 4$, $\zeta=100$, $k_{inter}= 10^{-3}$ and $Q=10$.
        }
	\FloatBarrier
        \label{fig:mechanism}
\end{figure}

The mixed-order nature of the spatial transition in the RR modular network (model (b)), is characterized by the critical exponent $\beta$, which governs how the size of the giant component vanishes as the system approaches the critical point. As shown in the inset of Fig. \ref{fig:comparing3models}, we measure $\beta=0.506$, in excellent agreement with this expectation, $\beta=1/2$, and with previous results for multiplex networks \cite{PhysRevLett.132.227401,PhysRevLett.105.048701,sallem2025interdependentnetworkstwointeractionsphysical,PhysRevE.93.042109}. 
The mixed-order phase transition is also characterized by an abrupt macroscopic collapse accompanied by critical dynamics and diverging time scales. To investigate whether the spatial transition in model (b) exhibits this behavior, we analyze the temporal dynamics of cascading failures. Specifically, we measure the duration of each cascade as the system approaches the spatial transition and observe a pronounced plateau, indicating long-lived near-critical dynamics (see Fig. \ref{fig:mechanism}). We quantify this behavior using the critical exponent 
$\zeta$, which characterizes the divergence of the cascade duration near criticality. As shown in Fig. \ref{fig:mechanism}, we obtain $\zeta=0.506$, in excellent agreement with the accepted value for systems undergoing mixed-order transitions. This agreement provides strong support that the spatial transition in model (b) belongs to the mixed-order universality class.

\section{Discussion}
In summary, we identify a modular–multiplex cascading mechanism that governs the two phase transitions observed in modular multiplex networks due to random failures. As the system nodes are progressively degraded, the first transition corresponds to the disconnection of modules, while the second marks the global collapse within the modules of the network. These transitions can be continuous, abrupt, or mixed-order, depending on the topology of the intra-module and the inter-module connectivity range. In particular, the abrupt transitions observed in the RR modular multiplex are accompanied by long-lived cascade plateaus and critical scaling, reflecting the coexistence of discontinuities and critical behavior.
A key factor underlying these distinct transition types is the range of the connectivity links between modules. In the RR modular multiplex, connectivity links are effectively long-range, enabling failures to propagate across the system without geometric constraints and resulting in an abrupt, first-order collapse. By contrast, when connectivity links between modules are short-range, see Fig. 1a, as in the lattice-based model, cascading failures are spatially constrained, leading to continuous or mixed-order transitions. We further conjecture that intermediate-range connectivity may induce a nucleation-driven collapse, in which localized damage nucleates compact regions of failure that subsequently spread and trigger system-wide breakdown.
The analytical determination of the module-disconnection threshold clarifies the origin of the first transition and demonstrates that it is controlled by intra-module connectivity and inter-module topology rather than by module size. From a broader perspective, these results emphasize that the range of connectivity links plays a central role in determining the universality class of cascading failures in modular multiplex systems, with direct implications for the resilience of real-world infrastructures.

\section{Acknowledgments}
This research is supported by grants from the Israel Science Foundation (Grant No. 201/25), the EU H2020 project DIT4TRAM, the EU H2020 Project OMINO (Grant No. 101086321), the VATAT National Foundation for climate and for power grids, the Israel Ministry of Innovation, Science $\&$ Technology (Grant No. 01017980) and the Germany-Israel project (DIP) grant.
	\FloatBarrier
\bibliographystyle{naturemag_4etal}
\bibliography{references}

\end{document}